\documentclass[aps,prl,showpacs,twocolumn,superscriptaddress]{revtex4-1} %,nofootinbib
\usepackage{graphicx}
\usepackage{bm}
\usepackage{amssymb} %maths
\usepackage{amsmath} %maths
\usepackage{amsthm}
\usepackage{hyperref}
\usepackage[utf8]{inputenc}
\hypersetup{
     colorlinks=true,      
    linkcolor=blue,        
    citecolor=blue,
    filecolor=magenta, 
    urlcolor=black          
}

\providecommand{\bra}[1]{\langle #1 \rvert}
\providecommand{\ket}[1]{\lvert #1 \rangle}
\providecommand{\braket}[2]{\langle #1 \rvert #2 \rangle}
\providecommand{\ketbra}[2]{\lvert  #1\rangle \langle #2 \rvert}
\newcommand{\expec}[1]{\langle #1\rangle}
\usepackage{mathbbol}

\begin{document}
\title{Entropic uncertainty relations from quantum designs}

\author{Andreas Ketterer}
\affiliation{Physikalisches Institut, Albert-Ludwigs-Universit\"at Freiburg, Hermann-Herder-Str. 3,
79104 Freiburg, Germany}
\affiliation{Naturwissenschaftlich-Technische Fakult\"at, Universit\"at Siegen, Walter-Flex-Str. 3, 57068 Siegen, Germany}
\author{Otfried G\"uhne}
\affiliation{Naturwissenschaftlich-Technische Fakult\"at, Universit\"at Siegen, Walter-Flex-Str. 3, 57068 Siegen, Germany}

\begin{abstract}

In the course of the last decades entropic uncertainty relations have attracted much attention not only due to their fundamental role as manifestation of non-classicality of quantum mechanics, but also as  major tools for applications of quantum information theory. Amongst the latter are protocols for the detection of quantum correlations or for the secure distribution of secret keys. In this work we  show how to derive entropic uncertainty relations for sets of measurements whose effects form quantum designs. The key property of quantum designs is their indistinguishability from truly random quantum processes as long as one is concerned with moments up to some finite order. Exploiting this characteristic enables us to evaluate polynomial functions of measurement probabilities which leads to lower bounds on sums of generalized entropies. As an application we use the derived uncertainty relations to investigate the incompatibility of sets of binary observables.

\end{abstract}

\maketitle 
%
%\section{Introduction}
\textit{Introduction.}--- Quantum mechanics prohibits observers to make simultaneous predictions about complementary properties of a physical system \cite{Messiah}.  This fundamental aspect manifests itself through restrictions of uncertainties of measurement outcomes performed on a number of identically prepared copies of a system and is strongly related to the noncommuting structure of the observables under consideration. The latter famously lead to the formulation of Heisenberg's uncertainty relation (UR) \cite{HeisenbergIneq} and Robertson's generalization thereof \cite{RobertsonIneq}. Uncertainty relations thus reflect the inherent nonclassical nature of quantum mechanics which makes them also central objects in the theory of quantum information \cite{EPRReview,SteveReviewUR}. 

The first URs have been formulated in terms of the variance of observables as a measure of their degree of uncertainty~\cite{HeisenbergIneq,RobertsonIneq}. However, it was recognized later on that also other uncertainty measures are able to capture the concept of complementarity and, in addition, come along with a number of advantageous other properties~\cite{WehnerEUR2}. For instance, the information theoretic interpretation of entropic uncertainty relations (EUR), which rely on entropies as quantifiers of uncertainty~\cite{WehnerEUR1,WehnerEUR2}, makes them ideal candidates for cryptographic protocols, such as quantum key-distribution or randomness generation \cite{CryptoReview}. Furthermore, EURs are an important building block for the formulation of criteria capable of detecting nonclassical properties of quantum mechancis, i.e.  entanglement~\cite{OtfriedReview,OtfriedEntEUR}, quantum steering~\cite{WalbornSteering1,WalbornSteering2,AnaSteeringEUR,ReviewSteeringCavalcanti,ReviewSteeringRoope,BrunnerEUR} or structures of measurement incompatibility~\cite{BrunnerEUR}. Importantly, these criteria are often applicable in discrete- as well as the continuous-variable regime which makes them a versatile tool \cite{DuanCrit,SimonCrit,WalbornSteering1,SteveReviewUR}.

The first EURs have been derived for sums of Shannon entropies \cite{ShannonEntropy} and they mostly involved small sets of observables, such as position and momentum \cite{BirluEUR}, or pairs of observables with finite spectrum \cite{DeutschEUR}. Famously, Maassen and Uffink derived the EUR~\cite{MaassenUffink}: 
 \begin{align}
S(\mathcal A)+S(\mathcal B)\geq \ln{(1/c)},\ \ \ c:=\max_{j,k}{|\braket{a_j}{b_k}|^2},
\label{eq:MaassenUffink}
 \end{align}
where $S(\mathcal A)$ ($S(\mathcal B)$) denotes Shannon's entropy of the probability distribution $\mathcal \{p^{(\mathcal A)}_k\}_{k}$  ($\mathcal \{p^{(\mathcal B)}_k\}_{k}$), originating from measurements of the observable $\mathcal A$ ($\mathcal B$) on a quantum state $\rho$, and where the $\ket{a_j}$'s ($\ket{b_k}$) denote the observable's eigenstates.

Since Maassen and Uffink there have been a number of generalizations of EURs which considered generalized entropy functions as well as larger sets of observables. For instance, a number of EURs for Shannon entropies have been generalized later on to the class of R\'enyi \cite{RenyiEntropy} and Tsallis \cite{TsallisEntropy} entropies \cite{MaassenUffink,ZoZorEUR}. Furthermore, by invoking the properties of observables with mutually unbiased eigenbases it was possible to derive of EURs involving more than two observables~\cite{MUBsReview}. Also the latter have been first obtained for Shannon entropies~\cite{PauliEURShannon,MolmerEUR}, and were later generalized to R\'enyi and Tsallis entropies \cite{WehnerMUBEUR,Rastegin1,Rastegin2,Rastegin3,JedEUR}.  However, deriving EURs for more observables than there exist MUBs for a given dimension is in general challenging.

In this work we take further steps in this direction and show how to derive EURs for more observables than the maximum number of existing MUBs. To do so, we exploit the pseudo-random properties of the effects representing the measurements under considerations. Whenever the latter form a so-called quantum design, we are able to prove  lower bounds on the sums of the respective generalized entropies. In this context, the property of being a quantum design can be seen as the natural generalization of the mutually unbiasedness of measurement effects, holding in particular for larger sets of observables. As an application of the derived EURs we use them to study the incompatibility properties the considered sets of binary observables.  

\textit{Quantum designs.}--- Let us denote the set of unit vectors in the finite-dimensional complex Hilbert space $\mathbb C^d$ as $\mathbb S^d$. Then, a {quantum $t$-design} is a set $\{\ket{\psi_k}\in\mathbb S^d\}_{k=1}^{K}$ which fulfills the property~\cite{Caves,EmersonQuDesign}:
\begin{align}
\frac{1}{K} \sum_{k=1}^{K} P_{t}(\ket{\psi_k}) =\int d\psi  P_{t}(\ket{\psi_k}), 
\label{eq:tDesignDef}
\end{align}
where $P_t:\mathbb S^d\rightarrow \mathbb C$ denotes a polynomial of degree at most $t$, and $d\psi$ the respective Haar measure on $\mathbb S^d$. 
Note that every $t$-design is also a $s$-design, with $s\leq t$. 

\begin{figure}[t]
\begin{center}
\includegraphics[width=0.47\textwidth]{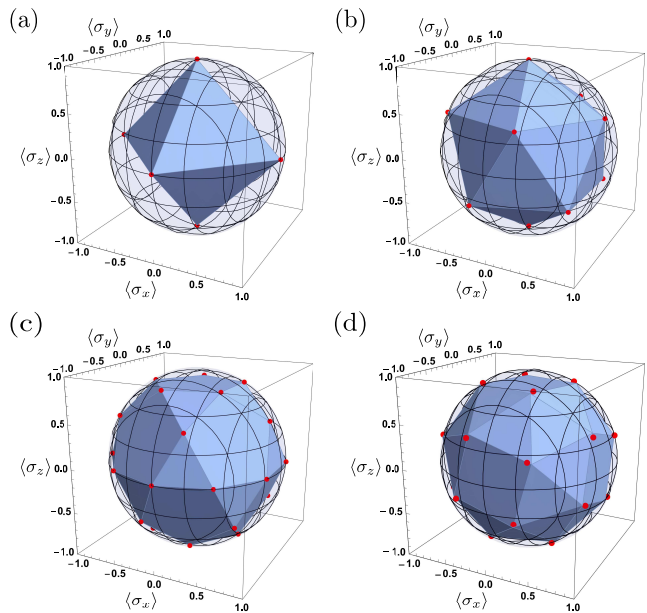}%
\end{center}
\caption{Plot of the Bloch vectors of various quantum designs, for $d=2$, and their corresponding polyhedra. The components of the Bloch vector are denoted as $\expec{\sigma_i}$, with the Pauli matrices $\sigma_x$, $\sigma_y$, and $\sigma_z$. (a)  $3$-design with $K_\text{octa}=6$ vertices forming an octahedron. (b) $5$-design with $K_\text{icosa}=12$ vertices forming an icosahedron. (c) $5$-design with $K_\text{icosi}=30$ vertices forming an icosidodekahedron. (d) $7$-design with $K_{7-\text{design}}=24$ vertices forming a deformed snub cube (red points). For comparison, a regular snub cube only forms a $3$-design (blue). } 
\label{fig_1}
\end{figure}

Quantum designs proved useful for a wide range of applications in quantum information theory \cite{ApplDesigns1,ApplDesigns2,ApplDesigns3,ApplDesigns4,ApplDesigns5,ApplDesigns6}. However, though their existence has been proven~\cite{existence}, the known examples of exact quantum designs are rare. While there is no general strategy that allows one to generate quantum designs with a given degree $t$, one can exploit group theoretical methods to find them in a number of relevant cases~\cite{ExamplesSphericalDesigns,Gross5Design}. In Fig.~\ref{fig_1} we present four examples of quantum designs in $\mathbb C^2$. In some of these cases quantum designs can be associated to polyhedra with different number of vertices. For instance, the $3$-design originating from the Clifford group (see Fig.~\ref{fig_1}(a)) forms an octahedron with $K_\text{octa}=6$ vertices, whereas the $5$-designs presented in Fig.~\ref{fig_1}(b) and (c) correspond to an icosahedron with $K_\text{icosa}=12$ and an icosidodekahedron with $K_\text{icosi}=30$ vertices, respectively. Lastly, the $7$-design with $K_{7-\text{design}}=24$ vertices shown in Fig.~\ref{fig_1}(d) represents a deformed snub cube, i.e., a regular snub cube with slightly smaller square faces and slightly larger triangular faces~\cite{ExamplesSphericalDesigns}.

An important property of quantum $t$-designs is that
\begin{align}
\sum_{k=1}^{K}\ketbra{\psi_k}{\psi_k}^{\otimes t}=K\mathcal D^{(t)}_d \mathbb 1_{t}^{\text{sym}},
\label{eq:DesignProjSym}
\end{align}
where $\mathbb 1_{t}^{\text{sym}}$ denotes the projector onto the symmetric subspace of $(\mathbb C^d)^{\otimes t}$, and $\mathcal D^{(t)}_d=t!(d-1)!/(t+d-1)!$ the inverse of its dimension~\cite{Caves}. Note that Eqs.~(\ref{eq:tDesignDef}) and (\ref{eq:DesignProjSym}) are even equivalent, a fact that leads to simple methods for the verification of quantum designs~\cite{Caves} and to the following Proposition. 

\begin{figure}[t!]
\begin{center}
\includegraphics[width=0.48\textwidth]{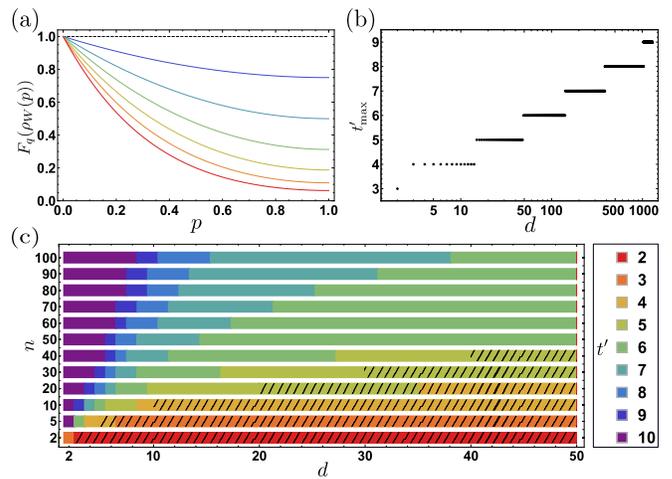}%
\end{center}
\caption{(a) Plot of $F_t[\tilde\rho(p)]$ as a function of the mixing parameter $p$ for different values $t=1$ (black, dashed) and $t=2,\ldots,7$ (blue to red). (b) Optimal value of $t'$, resulting from the maximization of the RHS of Eq.~(\ref{eq:EURRenyi1}) for projective measurements ($d=n$), as a function of the dimension $d$. (c) Optimal value of $t^\prime$, resulting from the maximization of Eq.~(\ref{eq:EURRenyi1}) for one rank-$1$ POVM ($M=1$), as a function of the number of outcomes $n=2,5,10,\ldots,100$ and the  dimension $d=1,\ldots,50$. The dashed area, i.e., values of $n$ and $d$ for which $n<d$, is excluded because we consider only rank-$1$ measurements. } 
\label{fig_2}
\end{figure}

\textit{Proposition 1:} Given a quantum $t$-design $\{\ket{\psi_k}\in\mathbb S^d\}_{k=1}^{K}$ and an arbitrary density matrix $\rho\in \mathbb C^{d\times d}$. Then,  Eq.~(\ref{eq:DesignProjSym})  implies
\begin{align}
\sum_{k=1}^K \bra{\psi_k}\rho\ket{\psi_k}^t = K \mathcal D^{(t)}_d F_t(\rho),
\label{eq:DesignBound}
\end{align}
where $F_t(\rho):=\text{tr}[\rho^{\otimes t}\mathbb 1_t^{\text{sym}}]$. \\
\indent We note that $F_t(\rho)$ can  be expressed as a sum of monomials of the moments $\mu_m=\text{tr}[\rho^m]$, with $m\leq t$, over the conjugacy classes of the symmetric group $S_t$ of permutations of $t$ elements. The latter are conveniently identified with the shape of Young diagrams or the partitions of $t$, denoted as $\lambda \vdash t$~\cite{AnaCombi}, leading to
\begin{align}
F_t(\rho):=\frac{1}{t!} \sum_{\lambda \vdash t}  h_\lambda \prod_{m^{(\lambda)}=1}^t \text{tr}[\rho^{m^{(\lambda)}}]^{k_m^{(\lambda)}}, 
\label{eq:Fnrho}
\end{align}
where $h_\lambda$ denotes the number of permutations contained in the conjugacy class $\lambda \vdash t$, the integers $m^{(\lambda)}$ the respective orders of the cycles associated with it, and $k_m^{(\lambda)}$ the number of times a cycle of order $m^{(\lambda)}$ occurs. As an example, for $t=2$, we find that $F_2(\rho)=(1+\text{tr}[\rho^2])/2$ is a linear function of the purity of $\rho$. For further details of the derivation of Eq.~(\ref{eq:Fnrho}) see Appendix~A.\\
\indent Concerning the properties of Eq.~(\ref{eq:Fnrho}), it is easy to see that $F_t(\rho)\leq 1$, with equality if and only if $\rho$ is a pure state, leading directly to a state independent upper bound of the LHS of Eq.~(\ref{eq:DesignBound}). Furthermore, as $\mu_m$ attains its minimum  $1/d^{m-1}$  for the maximally mixed state, we find that $F_t(\rho)$ is lower bounded by $[\Pi_{k=1}^{t-1}(k+d)]/t!d^{t-1}$. See Fig.~\ref{fig_2}(a) for an exemplary plot of $F_t[\tilde\rho(p)]$, where $\tilde\rho(p):=(1-p) \ketbra{0}{0}+p \mathbb 1/2$, with $0\leq p\leq 1$. It is evident that $F_t[\tilde\rho(p)]$ decreases monotonously towards the above minimum. 

\textit{EURs from quantum designs.}--- 
In the following we will focus on so-called R\'enyi and Tsallis entropies:
\begin{align}
H_\alpha(\{p_k\}_{k=1}^n)&=\frac{1}{1-\alpha}\ln{\left(\sum_{k=1}^n p_k^\alpha\right)},\label{eq:ReEnt}\\
T_q(\{p_k\}_{k=1}^n)&=\frac{1}{1-q}\left(\sum_{k=1}^n p_k^q-1\right), \label{eq:TsEnt}
\end{align}
with $\alpha,q \in \mathbb R_{>0}\backslash \{1\} $, which converge to the well-known Shannon entropy in the limit $\alpha,q \rightarrow 1$. The entropies~(\ref{eq:ReEnt}) and (\ref{eq:TsEnt}) are related through the monotonic function $f_r(x):=\ln{(1+(1-r)x)}/(1-r)$, with $r > 1$, which allows one to convert their respective values into each other. However, we emphasize that this is in general not possible for EURs in terms of the two entropies.  

As in Eq.~(\ref{eq:MaassenUffink}), we define the entropy of a positive operator valued measure (POVM) $\mathcal E$ consisting of a set of effects $\{E_k\}_{k=1}^n$, with $E_k\geq 0$ and $\sum_k E_k=\mathbb 1$, by applying Eqs.~(\ref{eq:ReEnt}) or (\ref{eq:TsEnt}) to the resulting outcome probability distribution $\{p^{(\mathcal E)}_k=\text{tr}[E_k\rho]\}_k$. In the particular case of a rank-$1$ measurement all the effects  are rank-$1$ operators of the form $E_k=d/n \ketbra{\psi_k}{\psi_k}$, with a projector $\ketbra{\psi_k}{\psi_k}$ onto the $1$-dimensional subspace defined by $\ket{\psi_k}$. 

With the resulting entropies as a measure for the degree of uncertainty of a measurement in hand we arrive at our  main result.

\textit{Theorem 1:} Given a quantum $t$-design $\{\ket{\psi_k}\in\mathbb S^d\}_{k=1}^{K}$ which originates from the effects of $M$ rank-$1$ POVMs $\{\mathcal B_m\}_{m=1}^M$, each having $n$ outcomes (i.e., $K=n M$). Then, we obtain the following  EUR for R\'enyi entropies:
\begin{align}
\sum_{m=1}^M H_\alpha (\mathcal B_m ) \geq \frac{M\alpha}{t^\prime(1-\alpha)}\ln{\left(\frac{d^{t^\prime}}{n^{(t^\prime-1)}} \mathcal D^{(t^\prime)}_d F_{t^\prime}(\rho) \right)},\label{eq:EURRenyi1}
\end{align}
for all $\alpha \geq t^\prime$, with $t'$ being an arbitrary integer in the interval $[2,t]$, and where $\mathcal D^{(t')}_d$ and $F_{t^\prime}(\rho)$ are defined in Proposition~1. Setting $F_{t^\prime}(\rho)$ equal to one yields the corresponding state independent version of Eq.~(\ref{eq:EURRenyi1}).

As each R\'enyi entropy in Eq.~(\ref{eq:EURRenyi1}) depends on the sum of the respective exponentiated outcome probabilities of the POVMs $\mathcal B_m$, we can use Proposition~1 in combination with the monotonicity of the $p$-norm $|\boldsymbol x|_p=(\sum_i x_i^p)^{1/p}$ and the function $f_\alpha(x):=\ln{(x)}/(1-\alpha)$, for $\alpha > 1$, to lower bound their sum and thus prove Theorem~1. For a detailed proof we refer the reader to Appendix~B and for an analog version of Theorem~1 involving Tsallis entropies to Appendix~C. Note that, due to the generality of Theorem~1, Eq.~(\ref{eq:EURRenyi1}) yields in particular EURs for sets of projective measurements $(n=d)$, and bounds on the entropy of single rank-$1$ POVMs with $n$ elements $(M=1)$~\cite{SingleEntropyPaper}. 

However, before we discuss concrete examples, the problem remains to find the optimal value of the parameter $t'$ in Eq.~(\ref{eq:EURRenyi1}). To do so, we  maximize the RHS of Eq.~(\ref{eq:EURRenyi1}) with respect to $t'$ for a given set of measurements. We discuss this maximization for the two above mentioned cases of projective measurements ($d=n$), and individual POVMs ($M=1$)~\cite{footnote1}. The result is presented in  Fig.~\ref{fig_2}(b) and (c), respectively. For projective measurements we find that, while for $d=2$ the best bound is obtained for $t^\prime=3$, for an increasing dimension larger values of $t^\prime$ yield better EUR bounds. In the second case, i.e., for individual POVMs, we find that large values of $t^\prime$ become useful for increasing number of outcomes $n$ depending on the dimension $d$. Also, we emphasize that the bounds of the state dependent version of Eq.~(\ref{eq:EURRenyi1}) become stronger for increasingly mixed states, as the minimum of $F_t(\rho)$ is attained for the maximally mixed one (see also Fig.~\ref{fig_2}(a)). 

Lastly, we want to stress that the existence of the EURs derived in Theorem~1 is conditioned on the existence of a set of observables having the desired design property. This is reminiscent of EURs for measurements in mutually  unbiased bases (MUB) \cite{MolmerEUR,Rastegin1,Rastegin2,Rastegin3}, which can be proven for any dimension while  complete sets of $d+1$ MUBs are not known for arbitrary $d$ \cite{ApplDesigns5}. Note that Eq.~(\ref{eq:EURRenyi1}) implies the bounds obtained for full sets of MUBs, as the latter were proven by exploiting the fact that MUBs form quantum $2$-designs~\cite{Rastegin1}. Hence, our procedure  generalizes the bounds obtained for MUBs to larger sets of observables.

\textit{Examples for qubit systems.}--- As concrete examples of the  introduced EURs we consider sets of qubit observables whose eigenstates $\ket{\nu_i^{(m)}}$, with $i=0,1$, form one of the quantum designs presented in Fig.~\ref{fig_1}. In case of the octahedron (Fig.~\ref{fig_1}(a)), Theorem~1 leads to the well-known EURs for the three MUBs, $\mathcal X=\{\ketbra{+}{+},\ketbra{-}{-}\}$, $\mathcal Y=\{\ketbra{+_i}{+_i},\ketbra{-_i}{-_i}\}$, and $\mathcal Z=\{\ketbra{0}{0},\ketbra{1}{1}\}$ \cite{Rastegin1,Rastegin2}. In particular, since MUBs form not only $2$-designs but also $3$-designs, we improve the R\'enyi version of these EURs for $\alpha \geq 3$ (see Appendix~C). Further on, for the icosahedron $5$-design, presented in Fig.~\ref{fig_1}(b), Theorem~1 yields the following EUR for a set of six qubit measurements $\{\mathcal B^{(\text{icosa})}_m\}_{m=1}^6$:
\begin{align}
\sum_{m=1}^6 H_\alpha (\mathcal B^{(\text{icosa})}_m ) &\geq   \frac{6\alpha}{t^\prime(1-\alpha)} \ln{\left(\frac{2 t^\prime!}{(1+t^\prime)!}\right)},\label{eq:EURRenyiQubit}
\end{align}
for all $\alpha,q \geq t'$, with an integer $1<t'\leq 5$. 

Equation~(\ref{eq:EURRenyiQubit}), together with its Tsallis counterpart, presented in Appendix~C, thus constitute two concrete examples of EURs involving a larger number of observables than the maximal number of MUBs for $d=2$. We present their bounds as a functions of $\alpha$ in Fig.~\ref{fig_3}(a) (see Appendix~C for a discussion of the Tsallis version of Eq.~(\ref{eq:EURRenyiQubit})). We see that the optimal bound of Eq.~(\ref{eq:EURRenyiQubit}) is reached for $t^\prime=2$ in the range $2\leq \alpha\leq 3$ and for $t^\prime=3$ when $\alpha \geq 3$. We emphasize that the EUR~(\ref{eq:EURRenyiQubit}) cannot be obtained by trivially combining EURs for two sets of measurements of MUB, as the condition $|\braket{\nu_i^{(m_1)}}{\nu_j^{(m_2)}}|=1/\sqrt{2}$ is not fulfilled for any combination of $i,j,m_1$ and $m_2$. This situation changes for a measurement constructed from the icosidodekahedron $5$-design (see Fig.~\ref{fig_1}(c)), which can be equally obtained by combining a set of five rotated MUBs.% 

 Lastly, we note that the $7$-design presented in Fig.~\ref{fig_1}(d) does not consist of pairs of mutually orthogonal states preventing us from constructing a set of projective qubits measurements from it. However, as discussed previously, Theorem~1 also provides bounds on the entropy of individual POVMs. In Fig.~\ref{fig_3}(b) we present these bounds for the POVM $\mathcal E^{(7\text{-design})}$ consisting of the mentioned $7$-design. As expected from Fig.~\ref{fig_2}(c), increasing values of $t^\prime$ yield better bounds for growing $\alpha$. However, as $t^\prime \leq 7$ one cannot reach the optimal value of $t^\prime=10$. 

\begin{figure}[t!]
\begin{center}
\includegraphics[width=0.49\textwidth]{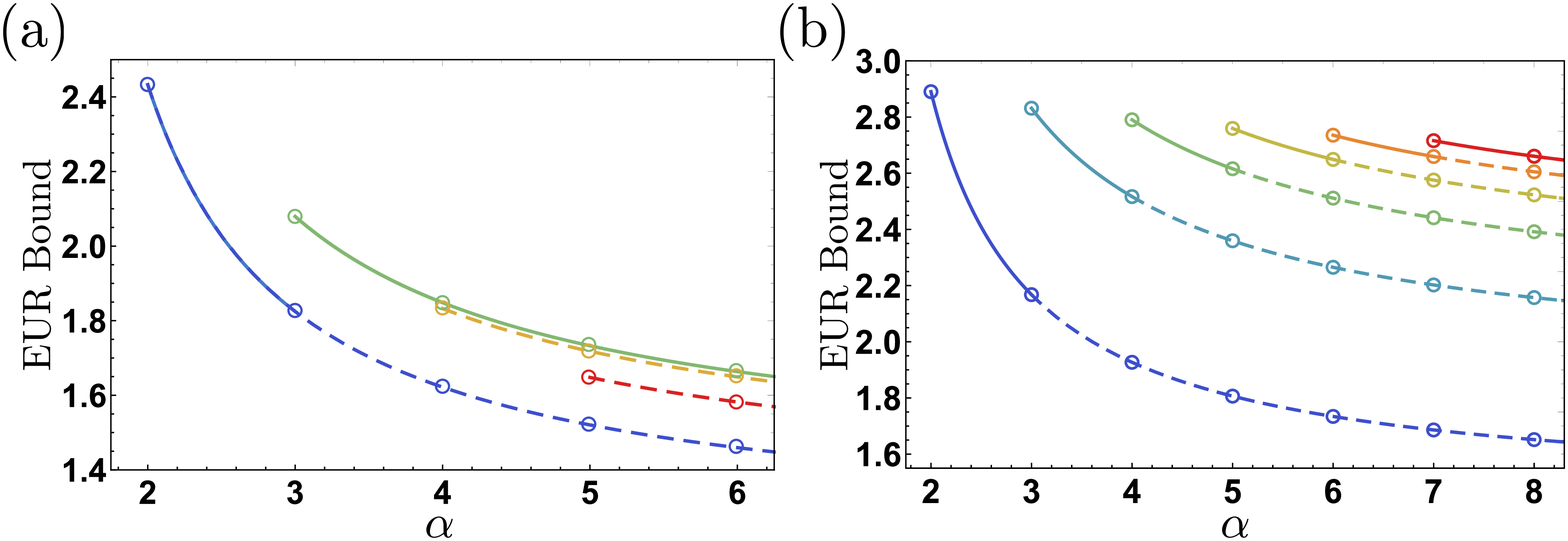}%
\end{center}
\caption{(a) Plot of the bound from Eq.~(\ref{eq:EURRenyiQubit}) as a function of the parameter $\alpha$ for different values $t^\prime=2$ (blue), $3$ (green), $4$ (yellow), and $5$ (red).  (b) Plot of the bound on the R\'enyi entropy of the individual POVM $\mathcal E^{(7\text{-design})}$, consisting of the $7$-design presented in Fig.~\ref{fig_1}(d), as a function of the parameter $q$ for different values $t^\prime=2$ (blue), $3$ (cyan), $4$ (green), $5$ (yellow), $6$ (orange) and $7$ (red). Solid lines represent the strongest bounds within a given range of $\alpha$ and $q$ and the circles indicate integer values.} 
\label{fig_3}
\end{figure}

\textit{Detection of measurement incompatibility.}--- As an application of the above derived EURs we use them to investigate the incompatibility properties of the previously discussed measurements. To do so, we adopt the notion of joint measurability which defines a set of measurements $\{\mathcal E_m\}_{m=1}^M$ to be compatible if its effects $E_{k|m}$ can be obtained via a classical post-processing of a larger joint measurement $\mathcal G=\{G_{\lambda}\}_\lambda$, i.e. $E_{k|m}=\sum_\lambda p(k|m,\lambda) G_\lambda$. If for a given set of measurements such a joint measurement does not exist we call them incompatible or not jointly measurable.

In order to investigate the incompatibility of a given set of measurements $\{\mathcal A_m\}_{m=1}^M$ we make use of the equivalence between steering and joint measurability problems~\cite{Roope1,Roope2}. More precisely, if two parties, Alice and Bob, share a pure maximally entangled state, the ability of Alice to steer Bob's state is equivalent to her using measurements which are not jointly measurable. Hence, we can exploit entropic steering inequalities (ESI) in order to test whether Alice's measurements are jointly measurable. Strategies for the construction of ESI based on EUR were presented in Refs.~\cite{BrunnerEUR,AnaSteeringEUR}. Applying the procedure of Ref.~\cite{BrunnerEUR} to the above derived EURs involving R\'enyi entropies leads to the ESI 
\begin{align}
\sum_{m=1}^M H_\alpha (\mathcal B_m|\mathcal A_m)\geq q(\{\mathcal B_m\}_{m=1}^M),
\label{eq:RenyiESI}
\end{align}
where $H_\alpha(X|Y)$ denotes the conditional R\'enyi entropy as defined in Refs.~\cite{CondRenyi1} (see also Appendix~D), and $q(\{\mathcal B\}_{m=1}^M)$ is the state independent bound of the corresponding R\'enyi EUR for Bob's measurements $\{\mathcal B\}_{m=1}^M$.

To test this procedure we regard the corresponding noisy counterparts $\mathcal A^{\eta}_m$ of Alice's measurements $\mathcal A_m=\{A_{k|m}\}$, defined as $A_{k|m}^\eta=\eta A_{k|m}+ (1-\eta) \mathbb 1/d$, with $0\leq \eta \leq 1$, and determine for which threshold value of $\eta$ Eq.~(\ref{eq:RenyiESI}) is violated. Starting of with the case of three MUBs, i.e. $\mathcal A^{(\text{Pauli})}=\{\mathcal X,\mathcal Y,\mathcal Z\}$, we recover that for $\alpha=2$ and $3$ a violation of Eq.~(\ref{eq:RenyiESI}) occurs for the well-known threshold $\eta\geq1/\sqrt{3}$ showing their triplewise incompatibility~\cite{StrucIncom}. Further on, we are able to apply the same procedure to the previously introduced set of  six icosahedron measurements $\{\mathcal B^{(\text{icosa})}_m\}_{m=1}^6$, which violate the ESI for $\eta\geq 1/\sqrt 3$, with $\alpha=2$ and $3$. Hence, in terms of the robustness against noise, the measurements $\{\mathcal B^{(\text{icosa})}_m\}_{m=1}^6$ are at least as incompatible as the set of Pauli measurements $\mathcal A=\{\mathcal X,\mathcal Y,\mathcal Z\}$ even though they are not mutually unbiased.

\textit{Conclusions.}--- We have proven bounds of entropic uncertainty relations in terms of R\'enyi and Tsallis entropies for measurements whose effects form quantum $t$-designs. To do so, we exploited the fact that quantum $t$-designs form projectors onto the symmetric subspace of a $t$-fold tensor product space which allowed us to make use of elements of representation theory of the symmetric group. The optimality of the introduced EURs was discussed for varying values of $t$ showing that larger degree designs can lead to stronger EUR bounds. Further on, we presented specific examples of these EURs, e.g., for measurements in mutually unbiased bases but also for larger sets of observables or for individual POVMs. Lastly, we probed the incompatibility properties of the introduced measurements through a set of steering inequalities which is based on the introduced EURs.

As an outlook one could imagine to  further generalize the introduced EUR for measurements whose effects form incomplete quantum designs, as it is possible to find EURs for less than $d+1$ MUBs in dimension $d$. Another interesting prospect is to derive EURs from approximate quantum designs, i.e., designs that fulfil the condition~(\ref{eq:tDesignDef}) up to some small error. Such a generalization would be intersting as there exist efficient constructions of approximate designs even for large values of $t$ \cite{ApproxDesign1,ApproxDesign2}.

\begin{acknowledgements}
We are indebted to Ren\'e Schwonnek, Timo Simnacher, Nikolai Wyderka and Xiao-Dong Yu for interesting discussions about entropic uncertainty relations. We acknowledge financial support from the ERC (Consolidator Grant 683107/TempoQ), and the DFG. AK acknowledges support from the Georg H. Endress foundation.
\end{acknowledgements}

%\appendix

%%%%%%%%%%%%%%%%%%%%%%%%%%%%%%%%%%%%%%%%%%%%%%%%%%%%%%%%%%%%%%%%%
\onecolumngrid
%%%%%%%%%%%%%%%%%%%%%%%%%%%%%%%%%%%%%%%%%%%%%%%%%%%%%%%%%%%%%%%%%
\vspace{8mm}
\begin{center}
{\large  APPENDIX}%ppendix} 
\end{center}
\section{A: Proof of Proposition 1}\label{app:A}
Here, we prove Proposition 1 which states that, given a quantum $t$-design $\{\ket{\psi_k}\in\mathbb S^d\}_{k=1}^{K}$ and an arbitrary density matrix $\rho\in \mathbb C^{d\times d}$, the following holds:
\begin{align}
\sum_{k=1}^K \bra{\psi_k}\rho\ket{\psi_k}^t = K \mathcal D^{(t)}_d F_t(\rho),
\label{app:DesignBound}
\end{align}
where $F_t(\rho):=\text{tr}[\rho^{\otimes t}\mathbb 1_t^{\text{sym}}]$ which can  be expressed as a sum of monomials of the moments $\mu_m=\text{tr}[\rho^m]$, with $m\leq t$, over the conjugacy classes of the symmetric group $S_t$ of permutations of $t$ elements. The latter are conveniently identified with the shape of Young diagrams or the partitions of $t$, denoted as $\lambda \vdash t$~\cite{AnaCombi}, leading to
\begin{align}
F_t(\rho):=\frac{1}{t!} \sum_{\lambda \vdash t}  h_\lambda \prod_{m^{(\lambda)}=1}^t \text{tr}[\rho^{m^{(\lambda)}}]^{k_m^{(\lambda)}}, 
\label{eq:Fnrho}
\end{align}
where $h_\lambda$ denotes the number of permutations contained in the conjugacy class $\lambda \vdash t$, the integers $m^{(\lambda)}$ the respective orders of the cycles associated with it, and $k_m^{(\lambda)}$ the number of times a cycle of order $m^{(\lambda)}$ occurs. 

\begin{proof}
To infer Eq.~(\ref{app:DesignBound}) we first note that any quantum $t$-design has the following property:
\begin{align}
\sum_{k=1}^{K}\ketbra{\psi_k}{\psi_k}^{\otimes t}=K\mathcal D^{(t)}_d \mathbb 1_{t}^{\text{sym}},
\label{app:DesignProjSym}
\end{align}
where $\mathbb 1_{t}^{\text{sym}}$ denotes the projector on the symmetric subspace $\text{Sym}^t(\mathbb C^d)$ of $(\mathbb C^d)^{\otimes t}$. By definition $\mathbb 1_{t}^{\text{sym}}$ can be expressed as a sum over permutation matrices $R_\pi$, which are defined as follows:
\begin{align}
R_\pi \ket{\psi_1}\otimes \ldots \otimes \ket{\psi_t}=\ket{\psi_{\pi^{-1}(1)}}\otimes \ldots \otimes \ket{\psi_{\pi^{-1}(t)}},
\label{app:DefRperm}
\end{align}
where $\pi$ denotes an element of the symmetric group $S_t$ which contains $t!$ elements, leading to the well-known expression 
\begin{align}
\mathbb 1_{t}^{\text{sym}}=\frac{1}{t!}\sum_{\pi \in S_t}R_\pi.
\end{align}
Now, by simply taking the expectation value of Eq.~(\ref{app:DesignProjSym}) with respect to a state $\rho^{\otimes t}$ we get
\begin{align}
\sum_{k=1}^K \bra{\psi_k}\rho\ket{\psi_k}^t = K \mathcal D^{(t)}_d \frac{1}{t!}\sum_{\pi\in S_t} \text{tr}[\rho^{\otimes t} R_\pi].
\label{app:DesignProjSym1}
\end{align}
Hence, it remains to be shown that $\sum_{\pi\in S_t}\text{tr}[\rho^{\otimes t} R_\pi]=\sum_{\lambda \vdash t} h_\lambda \prod_{m=1}^t \text{tr}[\rho^{m}]^{k^{(\lambda)}_m}$. 

To do so, we have to review some basics concerning the representation theory of the symmetric group $S_t$ of permutations of $t$ elements. First, note that $S_t$ contains $t!$ permutations each of which can be uniquely decomposed into a product of disjoint cyclic permutations (cycles)~\cite{RepTheo1}, also referred to as cycle decomposition. A cycle is of order $m$ if it acts on $m\leq t$ elements and the set of orders $\{m_1m_2\ldots \}$ of the cycles contained in a cycle decomposition, sorted in non-increasing order, determine the cycle type of a permutation. Alternatively, one can denote the cycle type of a permutation $\pi$ as $(1^{k_1}2^{k_2}\ldots t^{k_t})$, where $k_r$ is the number a cycle of order $r$ occurs. 

To make this clear we consider the following example. The permutation $\tilde\pi:\{12345678\}\rightarrow \{31254687\}$ of $S_8$ consists of the cycles $6\rightarrow 6$, $4\rightarrow 5\rightarrow 4$, $7\rightarrow 8\rightarrow 7$, and $2\rightarrow1\rightarrow3\rightarrow 2$, of order $1$, $2$, $2$, and $3$, respectively. The cycle decomposition of $\tilde\pi$ thus consists of $k_1=1$ cycles of order $1$, $k_2=2$ cycles of order $2$, and $k_3=1$ cycles of order $3$. The cycle type thus reads $(1223)$, or in the above introduced notation $(1^12^23^1)$, where cycles which occur $k_i=0$ times are not incorporated. 

It is evident that the cycle types of the permutations $\pi \in S_t$ can be identified with the  partitions of the integer $t$. A partition is a sequence of nonnegative integer numbers sorted in non-increasing order: $\lambda=(\lambda_1,\lambda_2,\ldots,\lambda_r,\ldots,\lambda_{l(\lambda)})$, of length $l(\lambda)$. Specifically,  a partition of an integer number $t$ is denoted by $\lambda \vdash t$, i.e. $t=\sum_i\lambda_i$, and the total number of partitions of an integer $t$ by $ P_t$. 
Furthermore, the cycle types also label the conjugacy classes of $S_t$, i.e. the equivalence classes of permutations that are conjugate to each other. Two permutations $\pi,\pi^\prime \in S_t$ are conjugates if it exists a third permutation $\sigma\in S_t$, s.t. $\pi=\sigma\pi^\prime \sigma^{-1}$. The number $h_\lambda$ of permutations contained in the conjugacy class $\lambda$ is then determined by the respective cycle type $(1^{k_1}2^{k_2}\ldots t^{k_t})$ through the formula~\cite{RepTheo1}:
\begin{align}
h_\lambda=\frac{t!}{k_1!1^{k_1}k_2!2^{k_2}\cdot\ldots\cdot k_t!t^{k_t}}.
\label{app:FormulaConCla}
\end{align}

As a consequence of the above discussion, we find that each permutation matrix $R_\pi$, corresponding to the permutation $\pi$, can be split into a product of cyclic permutation matrices $R_{m_1}\otimes R_{m_2}\otimes\ldots \otimes R_{m_r}$, where the $m_i$'s denote the orders of the involved cycles and $r$ the total number of cycles contained in the cycle decomposition of $\pi$. Note that each $R_{m_i}$ acts on a subspace $(\mathbb C^d)^{\otimes m_i}$ of $(\mathbb C^d)^{\otimes t}$, such that $\bigotimes_{i=1}^r (\mathbb C^d)^{\otimes m_i}=(\mathbb C^d)^{\otimes t}$.  If we take now the expectation value of $R_\pi$, where $\pi$ is a permutation of the conjugacy class $\lambda $ with cycle type $(1^{k^{(\lambda)}_1}2^{k^{(\lambda)}_2}\ldots t^{k^{(\lambda)}_t})$, with respect to the $t$-fold tensor product state $\rho^{\otimes t}$, we arrive at
\begin{align}
\text{tr}[\rho^{\otimes t} R_\pi]&=\text{tr}[\rho^{\otimes m_1 }R_{m_1}] \text{tr}[\rho^{\otimes m_2 }R_{m_2}] \ldots \text{tr}[\rho^{\otimes m_r }R_{m_r}] \label{app:CalcR1a}\\
&=\text{tr}[\rho R_{1}]^{k_1} \text{tr}[\rho^{\otimes 2 }R_{2}]^{k_2} \ldots \text{tr}[\rho^{\otimes t }R_{t}]^{k_t}\label{app:CalcR1b}\\
&=\text{tr}[\rho^{2}]^{k_2}\text{tr}[\rho^{3}]^{k_3} \ldots \text{tr}[\rho^{t }]^{k_t}. \label{app:CalcR1c}
\end{align}
In the second line~(\ref{app:CalcR1b}) we grouped together terms of the same order $m_i$, and in line~(\ref{app:CalcR1c}) we used that the expectation value of any cyclic permutation of order $m$ with respect to a state $\rho^{\otimes m}$ is equal to $\text{tr}[\rho^m]$, a fact that can be easily inferred from the definition~(\ref{app:DefRperm}) of the permutation matrices $R_\pi$. Lastly, summing over all permutations $\pi \in S_t$ yields
\begin{align}
\sum_{\pi \in S_t} \text{tr}[\rho^{\otimes t} R_\pi]&=\sum_{\pi \in S_t} \text{tr}[\rho^{2}]^{k^{(\lambda)}_2} \ldots \text{tr}[\rho^{t }]^{k^{(\lambda)}_t}\\
&=\sum_{\lambda \vdash t} h_\lambda \text{tr}[\rho^{2}]^{k^{(\lambda)}_2}\ldots \text{tr}[\rho^{t }]^{k^{(\lambda)}_t}\\
&=\sum_{\lambda \vdash t} h_\lambda \prod_{m=1}^t \text{tr}[\rho^{m}]^{k^{(\lambda)}_m},
\end{align}
where we used that the conjugacy classes of $S_t$ are labeled by the partitions $\lambda$ of $t$, and that each conjugacy class contains $h_\lambda$ elements (according to Eq.~(\ref{app:FormulaConCla})). This completes the proof of proof of Prop.~1.
\end{proof}

 In order to get familiar with the structure of the function $F_t(\rho)$ we discuss in the following some exemplary cases. Trivially, we find that for $t=1$ we have $F_1(\rho)=1$, as $S_1$ contains only a single element. In contrast, $S_2$ contains exactly two elements, i.e. the identity $id$ and  the transposition $\pi_{12}$, which are cycles of order $1$ and $2$, respectively. We thus find that 
 \begin{align}
F_2(\rho)&= \frac{1}{2} \left(1 + \text{tr}[\rho^2]\right).
 \end{align}
 For $t=3$ the situation gets slightly more complicated as $S_3$ has six elements. However, it is easy to see that those elements can be grouped into three conjugacy classes according to the three possible arrangements of Young diagrams which can be labeled by the partitions $(111)$, $(12)$ and $(3)$ of $t=3$. According to Eq.~(\ref{app:FormulaConCla}) we thus have that $h_{(111)}=1$, $h_{(12)}=3$ and $h_{(3)}=2$. Using the previously introduced alternative notation of the partitions $\lambda$ as cycle types we can write $(1^3)$, $(12)$ and $(3)$, thus inferring that 
 \begin{align}
F_3(\rho)&=\frac{1}{6} \left(1 + 3 \text{tr}[\rho^2] + 2 \text{tr}[\rho^3]\right). 
\end{align}
Following the same strategy for $t=4,5,6,7$ yields
\begin{align}
F_4(\rho)&=\frac{1}{24} \left(1+6 \text{tr}[\rho^2]+3 \text{tr}[\rho^2]^2+8 \text{tr}[\rho^3]+6 \text{tr}[\rho^4]\right), \\
F_5(\rho)&=\frac{1}{120} \left(1+15 \text{tr}[\rho^2]^2+20 \text{tr}[\rho^3] \text{tr}[\rho^2]+10 \text{tr}[\rho^2]+20 \text{tr}[\rho^3]+30 \text{tr}[\rho^4]+24 \text{tr}[\rho^5]\right),\\
F_6(\rho)&=\frac{1}{720} \left(1+15 \text{tr}[\rho^2]^3+45 \text{tr}[\rho^2]^2+120 \text{tr}[\rho^3] \text{tr}[\rho^2]+90 \text{tr}[\rho^4] \text{tr}[\rho^2]+15 \text{tr}[\rho^2]+40 \text{tr}[\rho^3]^2\right. \\
& \left.+40 \text{tr}[\rho^3]+90 \text{tr}[\rho^4]+144 \text{tr}[\rho^5]+120 \text{tr}[\rho^6]\right),\nonumber\\
F_7(\rho)&=\frac{1}{5040} \left(1 + 21 \text{tr}[\rho^2] + 
   105 \text{tr}[\rho^2]^2 + 105 \text{tr}[\rho^2]^3 + 
   70 \text{tr}[\rho^3] + 
   420 \text{tr}[\rho^2] \text{tr}[\rho^3] \right. \\
   &\left.+ 210 \text{tr}[\rho^2]^2 \text{tr}[\rho^3] + 
   280 \text{tr}[\rho^3]^2 + 210 \text{tr}[\rho^4] + 
   630 \text{tr}[\rho^2] \text{tr}[\rho^4] + 
   420 \text{tr}[\rho^3] \text{tr}[\rho^4] \right.  \nonumber \\
   &+\left. 
   504 \text{tr}[\rho^5] + 
   504 \text{tr}[\rho^2] \text{tr}[\rho^5] + 
   840 \text{tr}[\rho^6] + 720 \text{tr}[\rho^7]\right).\nonumber
 \end{align}
Lastly, we note that in Ref.~\cite{ElbenDesigns}  a similar formula has been derived in a different context using the properties of unitary designs. Unitary designs are the analogs of quantum designs on the level of unitary transformations, i.e., a set $\{U_k\in \mathcal U(d)|k=1,\ldots,K\}$ of unitary $(d\times d)$-matrices is called a unitary $t$-design if the following property holds:
\begin{align}
\frac{1}{K} \sum_{k=1}^{K} P_{t,t}(U) =\int dU  P_{t,t}(U_k), 
\label{eq:tDesignDef}
\end{align}
where $P_{t,t}:\mathcal U(d)\rightarrow \mathbb C$ denotes a polynomial of degree at most $t$ in the entries of the respective unitary matrix and their complex conjugates, and $dU$ is the Haar measure on $\mathcal U(d)$.

\section{B: Proof of Theorem 1}\label{app:B}
Next, we prove Theorem~1 which states that, given a quantum $t$-design $\{\ket{\psi_k}\in\mathbb S^d\}_{k=1}^{K}$ which originates from the effects of $M$ rank-$1$ POVMs $\{\mathcal B_m\}_{m=1}^M$, each having $n$ outcomes (i.e. $K=n M$). Then, we obtain the following  EUR for R\'enyi entropies:
\begin{align}
\sum_{m=1}^M H_\alpha (\mathcal B_m ) \geq \frac{M\alpha}{t^\prime(1-\alpha)}\ln{\left(\frac{d^{t^\prime}}{n^{(t^\prime-1)}} \mathcal D^{(t^\prime)}_d F_{t^\prime}(\rho) \right)},\label{app:EURRenyi1}
\end{align}
for all $\alpha \geq t^\prime$, with $t'$ being an arbitrary integer in the interval $[2,t]$, and where $\mathcal D^{(t^\prime)}_d$ and $F_{t^\prime}(\rho)$ are defined in Proposition~1. Setting $F_{t^\prime}(\rho)$ equal to one yields the corresponding state-independent version of Eq~(\ref{app:EURRenyi1}).

\begin{proof}
Evaluating the sum of entropies in Eq.~(\ref{app:EURRenyi1}) yields
\begin{align}
\sum_{m=1}^M H_\alpha(\mathcal B_m )&=\frac{1}{1-\alpha}\sum_{k=1}^M\ln{\left(\sum_{k=1}^n \text{tr}\left[\rho\frac{d}{n} \ketbra{b^{(m)}_k}{b^{(m)}_k}\right]^\alpha \right)}\label{app:RenyiProof1a}\\
&=\frac{1}{1-\alpha}\sum_{k=1}^M\ln{\left(\sum_{k=1}^n \left(\frac{d}{n}\right)^\alpha  \bra{b^{(m)}_k}\rho\ket{b^{(m)}_k}^\alpha \right)}.
\label{app:RenyiProof1b}
\end{align} 
Next, we use the monotonicity of the $p$-norm $|\boldsymbol x|_p$ which states that
\begin{align}
|\boldsymbol x|_q=\left(\sum_j x_j^q\right)^{1/q}\leq \left(\sum_j x_j^p\right)^{1/p}=|\boldsymbol x|_p
\label{app:pNorm}
\end{align}
for $1\leq p< q\leq\infty$, and apply it to the argument of the logarithm in Eq.~(\ref{app:RenyiProof1b}). This, in combination with the fact that the function $f_\alpha(x):=\ln{(x)}/(1-\alpha)$ is monotonically decreasing for $\alpha > 1$, allows us to estimate the RHS of Eq.~(\ref{app:RenyiProof1b}) and arrive at
\begin{align}
\sum_{m=1}^M H_\alpha(\mathcal B_m )
&\geq \frac{1}{1-\alpha}\sum_{k=1}^M\ln{\left(\left[\sum_{k=1}^n \left(\frac{d}{n}\right)^{t^\prime}  \bra{b^{(m)}_k}\rho\ket{b^{(m)}_k}^{t^\prime} \right]^{\alpha/{t^\prime}}\right)}\label{app:RenyiProof2a}\\
&= \frac{\alpha}{t^\prime(1-\alpha)}\sum_{k=1}^M\ln{\left(\sum_{k=1}^n \left(\frac{d}{n}\right)^{t^\prime}  \bra{b^{(m)}_k}\rho\ket{b^{(m)}_k}^{t^\prime} \right)}.
\label{app:RenyiProof2b}
\end{align}
Further on, we use the convexity of $f_\alpha(x)$, for $\alpha \geq 1$, and Jensen's inequality to get
\begin{align}
 \frac{\alpha}{t^\prime(1-\alpha)}\sum_{k=1}^M\ln{\left(\sum_{k=1}^n \left(\frac{d}{n}\right)^{t^\prime}  \bra{b^{(m)}_k}\rho\ket{b^{(m)}_k}^{t^\prime} \right)}\geq \frac{\alpha M}{t^\prime(1-\alpha)}\ln{\left(\frac{d^{t^\prime}}{M n^{t^\prime}}\sum_{k=1}^M\sum_{k=1}^n   \bra{b^{(m)}_k}\rho\ket{b^{(m)}_k}^{t^\prime} \right)}.
 \label{app:RenyiProof3}
\end{align}
Now, we can apply Proposition~1 according to which we have that
\begin{align}
\sum_{m=1}^M\sum_{k=1}^n \bra{b^{(m)}_k}\rho\ket{b^{(m)}_k}^{t^\prime} =n M \mathcal D^{(t^\prime)}_d F_{t^\prime}(\rho)\leq n M \mathcal D^{(t^\prime)}_d
\label{app:RenyiProof4}
\end{align}
for all $t^\prime \leq t$, and apply it to the Eq.~(\ref{app:RenyiProof3}) what finally yields
\begin{align}
\sum_{m=1}^M H_\alpha (\mathcal B_m ) \geq \frac{M\alpha}{t^\prime(1-\alpha)}\ln{\left(\frac{d^{t^\prime}}{n^{(t^\prime-1)}} \mathcal D^{(t^\prime)}_d F_{t^\prime}(\rho) \right)},\label{app:RenyiProof5}
\end{align}
for all $\alpha \geq t^\prime$, with an integer $1<t^\prime\leq t$. Lastly, imposing $F_{t^\prime}(\rho)\leq 1$, for all $\rho$, yields the corresponding state independent form of Eq.~(\ref{app:RenyiProof5}): 
\begin{align}
\sum_{m=1}^M H_\alpha (\mathcal B_m ) \geq \frac{M\alpha}{t^\prime(1-\alpha)}\ln{\left(\frac{d^{t^\prime}}{n^{(t^\prime-1)}} \mathcal D^{(t^\prime)}_d  \right)}.\label{app:RenyiProof6}
\end{align}
\end{proof}

The optimality of the bound in Eq.~(\ref{app:RenyiProof6}) as a function of $t^\prime$ has been discussed in Fig.~2(b) of the main text for the case of projective measurements and individual POVMs. We also discussed the following examples of Theorem~1. First, we reproduce the EURs for a complete set of MUBs which leads to
\begin{align}
\sum_{m=1}^{d+1} H_\alpha (\mathcal B_m ) \geq \frac{(d+1)\alpha}{t^\prime(1-\alpha)}\ln{\left(d\mathcal D^{(t^\prime)}_d  \right)}.\label{app:RenyiEURMUBs}
\end{align}
with $\alpha\geq t^\prime$ and $t^\prime\leq 3$. Note that Eq.~(\ref{app:RenyiEURMUBs}) reproduces the result obtained in Ref.~\cite{Rastegin1} for $t^\prime=2$ which is the optimal case in the interval $\alpha\in [2,3)$. However,  for $\alpha \geq 3$ we improve upon the result obtained in Ref.~\cite{Rastegin1} as we can set $t^\prime=3$ which is a consequence of the fact that MUBs not only form $2$-designs but also $3$-designs. In particular, for $d=2$ with the MUBs $\mathcal X=\{\ketbra{+}{+},\ketbra{-}{-}\}$, $\mathcal Y=\{\ketbra{+_i}{+_i},\ketbra{-_i}{-_i}\}$, and $\mathcal Z=\{\ketbra{0}{0},\ketbra{1}{1}\}$, we obtain
\begin{align}
H_\alpha (\mathcal X)+H_\alpha (\mathcal Y)+H_\alpha (\mathcal Z) \geq \left\{
                \begin{array}{lll}
                   \frac{3\alpha}{2(1-\alpha)}\ln{\left(2/3 \right)},&\text{for} &2\leq \alpha < 3 \\ \\
                  \frac{\alpha}{(1-\alpha)}\ln{\left(1/2  \right)}, &\text{for} &\alpha \geq 3.
                \end{array}
                \right.
                \label{app:EURRenyiQubitMUB}
\end{align}
Next, we discussed the EUR for the qubit measurements $\{\mathcal B^{(\text{icosa})}_m\}_{m=1}^6$ whose effects form an icosahedron:
\begin{align}
\sum_{m=1}^6 H_\alpha (\mathcal B^{(\text{icosa})}_m ) &\geq   \left\{
                \begin{array}{lll}
                  \frac{3\alpha}{1-\alpha} \ln\left(2/3\right), &\text{for} &2\leq \alpha < 3 \\ \\
                   \frac{2\alpha}{1-\alpha} \ln\left(1/2\right), &\text{for} &\alpha \geq 3.
                \end{array}
                \right.
\label{app:EURRenyiQubitIcosa}
\end{align}
Interestingly, the bounds of Eqs.~(\ref{app:EURRenyiQubitMUB}) and (\ref{app:EURRenyiQubitIcosa}) are proportional to each other, however, the measurements $\{\mathcal B^{(\text{icosa})}_m\}_{m=1}^6$ are not mutually unbiased. Hence, we cannot reproduce Eq.~(\ref{app:EURRenyiQubitIcosa}) by simply applying twice Eq.~(\ref{app:EURRenyiQubitMUB}). 

Furthermore, we also discussed bounds on entropies of individual POVMs whose effects form quantum designs. For instance, in case of the POVM $\mathcal E^{(7\text{-design})}=\{24/2 \ketbra{\kappa_i}{\kappa_i}\}_{i=1}^{24}$ made of the $24$ element improved snub cube $7$-design, presented in Fig.~1(d) of the main text, we find
\begin{align}
H_\alpha(\mathcal E^{(7\text{-design})})\geq \frac{\alpha}{(\alpha-1) t^\prime}\ln\left(\frac{t^\prime+1}{2^{3-2t'} \times 3^{1-t'}}\right),
\end{align}
for $\alpha \geq t^\prime$ and with $t^\prime \leq 7$. This bound has been discussed in Fig.~3(b) of the main text.

\section{C: Theorem for Tsallis entropies}\label{app:C}
We can  formulate an equivalent version of Theorem~1 for Tsallis entropies. 

\textit{Theorem~2} Given a quantum $t$-design $\{\ket{\psi_k}\in\mathbb S^d\}_{k=1}^{K}$ which originates from the effects of $M$ rank-$1$ POVMs $\{\mathcal B_m\}_{m=1}^M$, each having $n$ outcomes (i.e. $K=n M$). Then, we obtain the following  EUR  for R\'enyi entropies:
\begin{align}
\sum_{m=1}^M T_q (\mathcal B_m) &\geq \frac{1}{1-q}\left([dM\mathcal D^{(t^\prime)}_d F_{t^\prime}(\rho)]^{q/t^\prime}-M\right),\label{app:EURTsallis1}
\end{align}
for all $q \geq t^\prime$, with $t'$ being an arbitrary integer in the interval $[2,t]$, and where $\mathcal D^{(t')}_d$ and $F_{t^\prime}(\rho)$ are defined as in Proposition~1. Setting $F_{t^\prime}(\rho)$ equal to one yields the corresponding state independent version of Eq~(\ref{app:EURTsallis1}).

To prove Theorem~2 we first evaluate the sym of entropies in Eq.~(\ref{app:EURTsallis1}), yielding
\begin{align}
\sum_{m=1}^M T_q(\mathcal B_m )&=\frac{1}{1-q}\sum_{k=1}^M\left(\sum_{k=1}^n \text{tr}\left[\rho\frac{d}{n} \ketbra{b^{(m)}_k}{b^{(m)}_k}\right]^q -1 \right)\label{app:TsaProof1a}\\
&=\frac{1}{q-1}\left(M-\sum_{k=1}^M\sum_{k=1}^n \left(\frac{d}{n}\right)^\alpha  \bra{b^{(m)}_k}\rho\ket{b^{(m)}_k}^q  \right)\label{app:TsaProof1b}
\end{align} 
Next, we apply the monotonicity of the $p$-norm $|\boldsymbol x|_p$ to Eq.~(\ref{app:TsaProof1b}) to arrive at
\begin{align}
\sum_{m=1}^M T_q(\mathcal B_m )&\geq \frac{1}{q-1}\left(M-\left(\frac{d}{n}\right)^{q}\left(\sum_{k=1}^M\sum_{k=1}^n   \bra{b^{(m)}_k}\rho\ket{b^{(m)}_k}^{t^\prime} \right)^{q/t^\prime}  \right).\label{app:TsaProof2}
\end{align} 
Now, we use again Proposition~1 and the fact that $x^{q/t^\prime}$ is monotonically increasing with $x$, what yields
\begin{align}
\sum_{m=1}^M T_q (\mathcal B_m) &\geq \frac{1}{1-q}\left(\left[\frac{d^{t^\prime}}{n^{t^\prime-1}} M\mathcal D^{(t^\prime)}_d F_{t^\prime}(\rho)\right]^{q/t^\prime}-M\right),\label{app:TsaProof3}
\end{align}
for all $q\geq t^\prime$, with an integer $1<t^\prime \leq t$. Imposing $F_{t^\prime}(\rho)\leq 1$, for all $\rho$, again yields the state independent version
\begin{align}
\sum_{m=1}^M T_q (\mathcal B_m) \geq \frac{1}{1-q}\left(\left[\frac{d^{t^\prime}}{n^{t^\prime-1}} M\mathcal D^{(t^\prime)}_d \right]^{q/t^\prime}-M\right).
\label{app:TsaProof4}
\end{align}

\begin{figure}[t!]
\begin{center}
\includegraphics[width=0.7\textwidth]{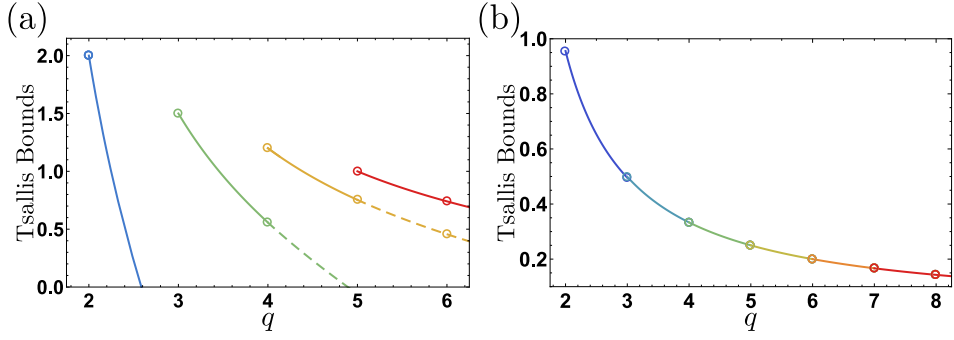}%
\end{center}
\caption{(a) Plot of the bound of the EUR~(\ref{app:EURTsallisQubit}) as a function of the parameter $\alpha$ for different values $t^\prime=2$(blue), $3$(green), $4$(yellow), and $5$(red).  (b) Plot of the bound on the Tsallis entropy of the individual POVM $\mathcal E^{(7\text{-design})}$, consisting of the $7$-design presented in Fig.~1(d) of the main text, as a function of the parameter $q$ for different values $t^\prime=2$ (blue), $3$ (cyan), $4$ (green), $5$ (yellow), $6$ (orange) and $7$ (red). Solid lines indicate for which ranges of $\alpha$ and $q$ the respective bounds are strongest and the circles emphasize integer values. } 
\label{app:fig_1}
\end{figure}
In particular, for $d=2$ with the MUBs $\mathcal X=\{\ketbra{+}{+},\ketbra{-}{-}\}$, $\mathcal Y=\{\ketbra{+_i}{+_i},\ketbra{-_i}{-_i}\}$, and $\mathcal Z=\{\ketbra{0}{0},\ketbra{1}{1}\}$, we obtain
\begin{align}
T_q (\mathcal X)+T_q (\mathcal Y)+T_q (\mathcal Z) \geq \left\{
                \begin{array}{lll}
                   \frac{1}{1-q}\left(2^{q/2}-3\right) ,&\text{for} &2\leq q < 3 \\ \\
                  \frac{1}{1-q} \left((3/2)^{q/3}-3\right), &\text{for} &q \geq 3.
                \end{array}
                \right.
                \label{app:EURTsallisQubitMUB}
\end{align}
Next, we discuss the EUR for the qubit measurements $\{\mathcal B^{(\text{icosa})}_m\}_{m=1}^6$ whose effects form an icosahedron:
\begin{align}
\sum_{m=1}^6 T_q (\mathcal I_m) &\geq \frac{1}{1-q}\left(\left[\frac{12 t^\prime!}{(t^\prime+1)!}\right]^{q/t^\prime}-6\right),\label{app:EURTsallisQubit}
\end{align}
for which the optimal values of $t^\prime$ are presented in Fig.~\ref{app:fig_1}(a) of the main text.

Furthermore, we also discuss bounds on entropies of individual POVMs whose effects form quantum designs. For instance, in case of the POVM $\mathcal E^{(7\text{-design})}=\{24/2 \ketbra{\kappa_i}{\kappa_i}\}_{i=1}^{24}$ made of the $24$ element improved snub cube $7$-design, presented in Fig.~1(d) of the main text, we find
\begin{align}
T_q(\mathcal E^{(7\text{-design})})\geq \frac{1}{1-q}\left(2^{q/t^\prime} \left(\frac{2^{3-2 t'}\times 3^{1-t'}}{t'+1}\right)^{q/t^\prime}-1\right),
\label{app:EURTsallisPOVM}
\end{align}
for $q \geq t^\prime$ and with $t^\prime \leq 7$. In Fig.~\ref{app:fig_1}(b) we present the resulting bounds of Eq.~(\ref{app:EURTsallisPOVM}) as a function of $q$ for different values of $t^\prime$. In the latter case varying $t^\prime$ has a very little influence on the value of the bound. Hence, $t^\prime=2$ can be assumed for all values of $q$.

\section{D: Entropic steering inequalities}\label{app:C}

There have been several works devoted to the derivation of quantum steering inequalities from entropic uncertainty relations~\cite{WalbornSteering1,WalbornSteering2,BrunnerEUR,AnaSteeringEUR}. These derivations were first based on EURs involving Shannon entropies~\cite{WalbornSteering1,WalbornSteering2}, but have been generalized to Tsallis and R\'enyi entropies later on~\cite{AnaSteeringEUR,BrunnerEUR}. In the following, we focus on the approach presented in Ref.~\cite{BrunnerEUR} as it is the only one applicable to EURs involving R\'enyi entropies for arbitrary $\alpha$. 

The main result of Ref.~\cite{BrunnerEUR} is a one-to-one mapping between any state independent EURs of the form 
\begin{align}
\sum_m H_m(\mathcal A_m)\geq q(\{\mathcal A_m\}),
\end{align}
with entropies $H_m$ and state independent bound $q(\{\mathcal A_m\})$, and entropic steering inequalities (ESI) of the form
\begin{align}
\sum_{m=1}^M H_m (\mathcal B_m|\mathcal A_m)\geq q(\{\mathcal B_m\}_{m=1}^M),
\label{eq:RenyiESI}
\end{align}
where $H_m(\mathcal B_m|\mathcal A_m)$ denotes the respective conditional entropy (see Eq.~(\ref{app:DefCondReEnt}) for a definition), provided the following two conditions hold:
\begin{itemize}
\item[(i)] The considered EUR holds true when conditioned on any classical random variable $\mathcal Y$,
\begin{align}
\sum_m H_m(\mathcal A_m|\mathcal Y)\geq q(\{\mathcal A_m\}).
\end{align}
\item[(ii)] The considered entropies are non-increasing under conditioning on additional information, i.e., 
\begin{align}
H_m(\mathcal X|\mathcal Y_1)\geq H_m(\mathcal X|\mathcal Y_1,\mathcal Y_2).
\end{align}
\end{itemize}

Here, we make use of this formalism to derive ESIs from EURs for R\'enyi entropies $H_\alpha$, with $\alpha \geq 2$. The associated conditional R\'enyi entropies are defined as follows:
\begin{align}
H_\alpha(\mathcal P_1|\mathcal P_2)=\frac{\alpha}{1-\alpha}\ln{\left(\sum_y p_y^{(2)} |\boldsymbol p^{(1|2)}_{\boldsymbol x| y}|_\alpha \right)},
\label{app:DefCondReEnt}
\end{align}
where $|\boldsymbol x|_\alpha$ denotes the $\alpha$-norm and $\boldsymbol p^{(1|2)}_{\boldsymbol x| y}$ is a vector containing conditional probabilities $p^{(1|2)}_{x| y}$ for a fixed $y$. 
Conditions (i) and (ii) have been proven in Refs.~\cite{BrunnerEUR} and \cite{CondRenyi1}, respectively. In this respect, the authors of Ref.~\cite{BrunnerEUR} focused on EURs involving at most two entropies, however, a generalization of the presented argument to more measurements is straightforward. All in all, this leads us to the following ESI presented in Eq.~(9) of the main text:
\begin{align}
\sum_{m=1}^M H_\alpha (\mathcal B_m|\mathcal A_m)\geq q(\{\mathcal B_m\}_{m=1}^M),
\label{app:RenyiESI}
\end{align}
where $H_\alpha(X|Y)$ denotes the conditional R\'enyi entropy as defined in Eq.~(\ref{app:DefCondReEnt}), and $q(\{\mathcal B_m\}_{m=1}^M)$ denotes the bound of the corresponding state independent R\'enyi EUR for Bob's measurements $\{\mathcal B_m\}_{m=1}^M$.

%%%%%%%%%%%%%%%%%%%%%%%%%%%%%%%%%%%%%%%%%%%%%%%%%%%%%%%%%%%%%%%%%
%\twocolumngrid
%%%%%%%%%%%%%%%%%%%%%%%%%%%%%%%%%%%%%%%%%%%%%%%%%%%%%%%%%%%%%%%%%


\begin{thebibliography}{99}
\bibitem{footnote1} Note that the pre-factor $\alpha/(\alpha-1)$ is monotonously decreasing, for $\alpha>1$, and its maximum is thus always obtained for $\alpha=t^\prime$.
\end{thebibliography}
\end{document}